\documentclass[letterpaper,10pt,twoside,twocolumn,final,notitlepage]{biophys}
\usepackage{helvet,times}
\usepackage{bm,textcomp}

\jno{kxl014} 
\gridframe{N}
\cropmark{N}

\doi{doi:  10.1016/S0006-3495(03)70021-6}
\fpage{2943}
\lpage{2956}
\volume{84}
\issue{5}
\Month{May}
\Year{2003}

\usepackage[round,numbers,sort&compress]{natbib}

\usepackage{graphicx}
\usepackage{amsmath}
\usepackage{amssymb}

\newcommand{\nl}{_{n\lambda}}

\newcommand{\snl}{{\sum_{\substack{n=0..4\\ \lambda=v,o}}}}
\newcommand{\sn}{{\sum_{n=0}^4}}
\newcommand{\re}{_\text{Ref}}

\begin{document}

\title{Perfect and near perfect adaptation in a model of bacterial chemotaxis}
\author{Bernardo A. Mello$^{*\dag}$ and Yuhai Tu$^*$}
\address{$^*$IBM T.J. Watson Research Center, P.O. Box 218, Yorktown Heights, NY 10598; 
and $^\dag$Physics Department, Catholic University of Brasilia, 72030-170, Brasilia, DF, Brazil}
\date{\today}

\begin{abstract}%
{The signaling apparatus mediating
bacterial chemotaxis can adapt to a wide range of persistent
external stimuli. In many cases, the bacterial activity returns to
its pre-stimulus level exactly and this ``perfect adaptability" is
robust against variations in various chemotaxis protein
concentrations. We model the bacterial chemotaxis signaling
pathway, from ligand binding to CheY phosphorylation. By solving
the steady-state equations of the model analytically, we derive a
full set of conditions for the system to achieve perfect
adaptation. The conditions related to the phosphorylation part of
the pathway are discovered for the first time, while other
conditions are generalization of the ones found in previous works.
Sensitivity of the perfect adaptation is evaluated by perturbing
these conditions. We find that, even in the absence of some of the
perfect adaptation conditions, adaptation can be achieved with
near perfect precision as a result of the separation of scales in
both chemotaxis protein concentrations and reaction rates, or
specific properties of the receptor distribution in different
methylation states. Since near perfect adaptation can be found in
much larger regions of the parameter space than that defined by
the perfect adaptation conditions, their existence is essential to
understand robustness in bacterial chemotaxis.}
{Preprint produced by the authors}
{}
\end{abstract}

\maketitle

\section*{INTRODUCTION}

The motion of coliform bacteria (such as {\em E-coli}) is driven
by rotation of several flagella attached to the cell body. When
the flagella rotate counter-clockwise (CCW), the flagella form a
bundle that pushes the bacterium in a smooth motion (runs) with a
high degree of directionality. On the other hand, when the
flagella rotate clockwise (CW), the flagella bundle flies apart
and the bacterium tumbles, randomizing the direction of the
subsequent run. The frequency with which the tumbling motion
occurs decreases with increasing concentration of attractant (or
decreasing concentration of repellent). As the result, the
bacterium performs a biased random walk towards higher
concentration of attractant. This mechanism gives the bacterium
its ability to follow the gradient of chemical concentration,
i.e., chemotaxis.

From the sensing of external stimulus to the activation of motor
regulator protein, a series of chemical reactions are involved in
relaying and regulating the signal. For recent review on bacterial
chemotaxis signaling pathway, see references (Falke, 1997; Bren,
2000; Bourret, 2002). The major players in the chemotaxis signal
transduction pathway are the transmembrane chemotaxis receptors
and 6 cytosolic proteins: CheA, CheB, CheR, CheW, CheY and CheZ.
The receptor forms a complex with the histidine kinase CheA
through the adaptor protein CheW. The receptor has a ligand
binding domain located at the periplasm to sense the external
signal, such as the concentration of attractant (or repellent).
The activity of CheA is affected by the properties of the
receptor, for example, whether the receptor is ligand bound or
not. When chemoattractant binds to receptor, CheA activity is
suppressed. The histidine kinase CheA, once activated, acquires a
phosphate group through autophosphorylation, and subsequently
transfers the phosphate group to the response regulator protein
CheY or the demethylation enzyme CheB. The phosphorylated CheY
(CheY-P) then interacts with the motor and increases the motor's
CW rotation bias. This is the ``linear" signal transfer part of
the bacterial chemotaxis pathway. Like many other biological
sensory systems, the bacterial chemotaxis pathway also has the
ability to adapt to persistent external stimulus. The adaptation
in bacterial chemotaxis is facilitated by the methylation and
demethylation of the receptor, which serves as the feedback
control of the system. The methylation and demethylation processes
are catalyzed by CheR and CheB-P respectively and are slow in
comparison with the other reactions.

Because of the excellent understanding of each individual reaction
of the pathway, mathematical modelling of bacterial chemotaxis
signal transduction has been very fruitful (Bray, 1993; Hauri,
1995; Barkai, 1997; Spiro, 1997; Morton-Firth, 1998; Morton-Firth,
1999; Yi, 2000). Besides being useful in understanding specific
aspects of chemotaxis experiments, modelling is essential in
gaining insight about general properties of biochemical networks.
One important general problem is to understand the functional
stability of biochemical networks under changes of various pathway
parameters, such as concentrations of enzymes and reaction rates.
Parameter fluctuations are inherent for biological systems in the
real world, so robustness, i.e., the insensitivity of important
system properties with respect to parameter variation and
fluctuation of protein concentrations, is crucial for the proper
functioning of the biological systems.

Experimentally, it was observed that after initial response to
some external stimulus, such as sudden changes of aspartate
concentrations, the bacteria tumbling frequency often reverts to
its original value with high accuracy, independent of the strength
of the external stimulus (Berg, 1972). This accurate adaptation is
generally believed to contribute to the high sensitivity of
bacterial chemotaxis to a wide range of external stimulus (5
orders of magnitude). In a recent work, Barkai and Leibler
(Barkai, 1997) investigated the robustness of perfect adaptation
in bacterial chemotaxis, they used a two-state (active or
inactive) model (Asukura, 1984) for the receptor complex in
explaining the phenomena. In their model, they assumed that CheB
only demethylates active receptors, whereas CheR methylates all
receptors indiscriminately. They showed, by extensive simulation
of the two-state model, that as long as the above conditions are
satisfied, adaptation is achieved with high precision, independent
of specific values of the rate constants or enzyme concentrations.
In a subsequent study, Alon {\it et al.} (1999) provided
experimental evidence for the robustness of the perfect adaptation
over large variations in chemotactic protein concentrations.

The Barkai-Leibler (BL) model clearly captured one of the
essential ingredients for perfect adaptation in bacterial
chemotaxis. Recently, Yi, Huang, Simon and Doyle (Yi, 2000)
further studied the Barkai-Leibler model analytically, and
summarized all the conditions for perfect adaptation within the BL
model beyond those identified in the original paper. However, the
BL model is a simplified description of the real chemotaxis
pathway. For example, the BL model neglects the phosphorylation
part of the pathway altogether and assumes the saturation of
methylation enzyme CheR, which is questionable (Morton-Firth,
1999).

In this work, we study a more complete model of the chemotaxis
signal transduction pathway, similar to the deterministic version
of the model proposed by Morton-Firth {\it et al.} (1998), where
both the methylation and phosphorylation processes are taken into
account. Our goals are to understand whether (mathematically)
perfect adaptation, defined as when steady-state CheY-P
concentration is independent of ligand concentration, can be
achieved for the full model, and to identify the conditions for
such perfect adaptation. The sensitivity of the perfect
adaptability, or robustness, is then studied by perturbing these
conditions. Such study can help us understand adaptation in real
biological systems where not all the perfect adaptation conditions
are satisfied, it can also provide possible explanations for cases
where perfect adaptation is not achieved, e. g., for serine
response (Berg, 1972).

\section*{MODEL}

\begin{table}
\caption{Chemical species and subspecies. Total concentrations are
taken from reference (Morton-Firth, 1999).} \label{Tab:species}
\begin{tabular}{llr}
\hline
\multicolumn{1}{l}{Species}&Description&Concentration\\
\hline
$[T^T]$ & Total taxis aspartate  & $2.5\,\mu\text{M}$\\
		& receptor (Tar)\\
$[T\nl]$& Receptor with $n$ methyl groups,\\
        & ligand binding site occupied \\
        & ($\lambda=o$) or vacant ($\lambda=v$) \\
$[T^F]$ & Free (CheR and CheB\\
		& unbound) receptor \\
$[T^P]$ & Phosphorylated receptor\\
$[T^U]$ & Unphosphorylated receptor\\
$[R^T]$ & CheR & $0.176\,\mu\text{M}$\\
$[R^F]$ & Free (not bound to $T$) CheR\\
$[B^T]$ & CheB & $2.27\,\mu\text{M}$\\
$[B^F]$ & Free (not bound to $T$) CheB\\
$[B^P]$ & Phosphorylated CheB\\
$[B^{PF}]$ & Free phosphorylated CheB\\
$[Y^T]$ & CheY & $18\,\mu\text{M}$\\
$[Y^P]$ & Phosphorylated CheY\\
\hline
\end{tabular}
\end{table}

For the purpose of this study, we consider only those receptors
that form complex with CheW and CheA. We label the receptor
complex by $T_{n\lambda}$, where $n(\in[0,4])$ is the number of
methyl groups added to the receptor and $\lambda$ $(=o,v)$
represents the ligand occupied ($o$) and vacant ($v$) state of the
receptor. Superscripts are also used to describe whether the
receptor complex is phosphorylated ($P$) or un-phosphorylated
($U$), bound to CheR/CheB-P or free ($F$). Superscript ($T$) is
used to label total concentrations of different proteins. The
superscripts are not mutually exclusive, e. g., $[B^{PF}]$ is the
concentration of phosphorylated free (not bound to receptor) CheB.
In table \ref{Tab:species}, some of the chemical species of the
chemotaxis pathway are shown, where the values of the total
concentrations are taken from (Morton-Firth, 1999), except for the
total CheR concentration, which we have reduced slightly in order
to have the same average methylation level as reported in
(Morton-Firth, 1999), where receptors other than Tar were included
in the simulation.

\begin{table*}
\caption{Chemotaxis signal transduction reactions}
\begin{center}

\label{Tab:reactions}
\begin{tabular}{lll}
\hline
Ligand binding & $T_{nv}+L\leftrightarrow T_nL(\equiv T_{no})$\\
\hline Methylation & $T_n+R^F \leftrightarrow T_nR$
& $T_nR\rightarrow T_{n+1}+R^F$\\
& $T_n+B^{PF} \leftrightarrow T_nB^P$
& $T_nB^{P}\rightarrow T_{n-1}+B^{PF}$\\
\hline
Phosphorylation & $T_n^{U} \rightarrow T_n^P$\\
& $T_n^P+Y^{U}\rightarrow T_n^{U}+Y^P$ & $Y^P \rightarrow Y^U$ \\
& $T_n^P+B^{UF}\rightarrow T_n^{U}+B^{PF}$ & $B^{PF} \rightarrow
B^{UF}$\\
\hline
\end{tabular}
\end{center}
\end{table*}

The bacterial chemotaxis pathway can be divided into 3 processes:
receptor ligand binding, receptor methylation/demethylation and
phosphorylation of CheA, CheB and CheY. The reactions involved in
each of the three processes are listed in table
\ref{Tab:reactions}. Since the ligand binding process is much
faster than the other two, the ligand binding reaction can be
considered to be always in quasi-equilibrium. The receptor's
ligand binding status directly affects both the CheA
auto-phosphorylation rate and the receptor
methylation/demethylation rates. The CheA auto-phosphorylation
rate is also affected by the methylation state of the receptor.
Finally, since only the phosphorylated CheB can efficiently
demethylate the receptor, the methylation process is also affected
by the phosphorylation process.

Some conformational change of the receptor complex is probably
responsible for the signaling from binding of ligand to
methylation and phosphorylation of the receptor complex (Bren,
2000;  Falke, 1997; Liu, 1997). The two-state model proposes that
the receptor complex has two states, active and inactive, with
only the active state capable of auto-phosphorylation. For a
receptor with $n$ methyl groups and a ligand occupancy status
described by $\lambda$ (vacant, $v$, or occupied, $o$), the
probability of being active is denoted by $P_{n\lambda}$. However,
there has been no direct experimental evidence in support of the
two-state model (Yi, 2000). More generally, $0\le P\nl\le 1$ can
be simply understood as the relative receptor activity for
receptor $T_{\nl}$ and the CheA auto-phosphorylation rate is
proportional to $P\nl$:
\begin{equation}
k^{P}_{n\lambda}=k^{P}P_{n\lambda},
\end{equation}
where $k^{P}$ is a constant independent of $n$ and $\lambda$.

\begin{figure}
\vspace{-1.5cm}
\hspace{-1cm}
\includegraphics[width=8.5cm,angle=270]{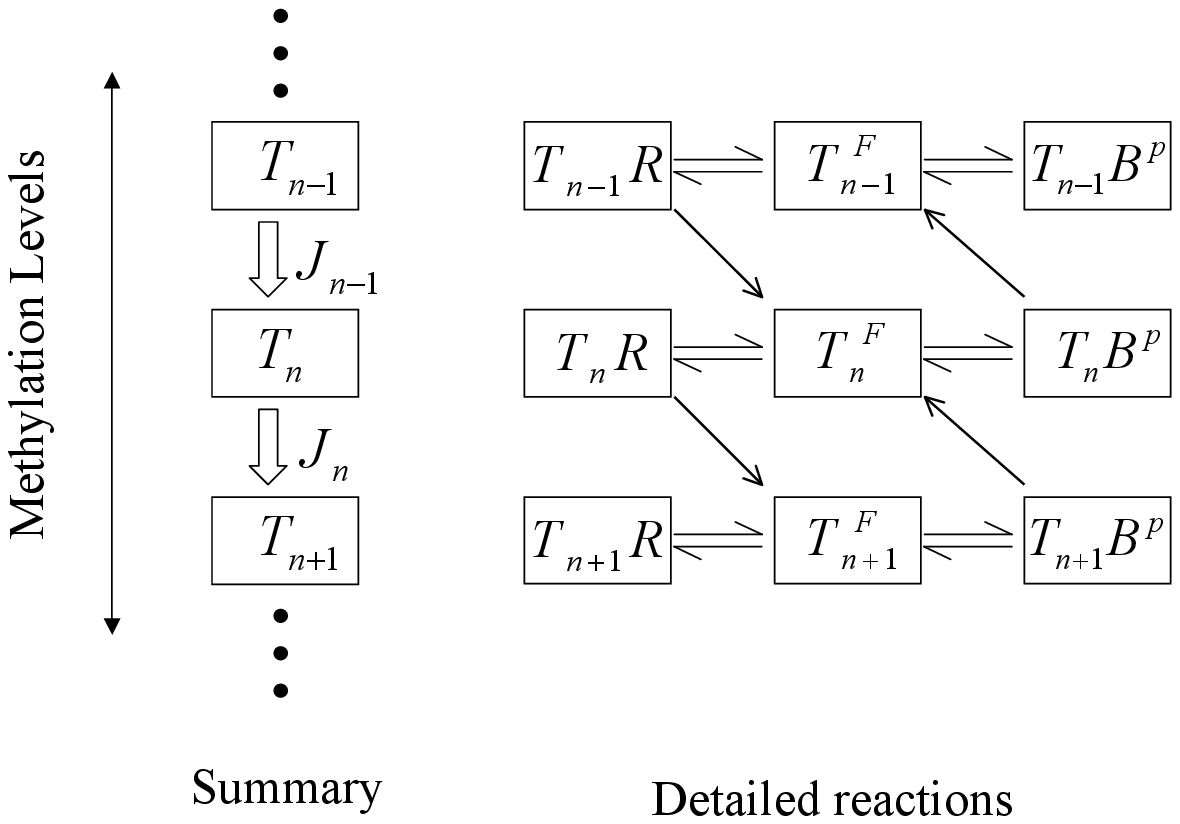}
\vspace{-1cm}
\caption{Illustration of the methylation and demethylation
reaction network, $n$ is the methylation level of the receptor
.}\label{mflow}
\end{figure}

In the following, we write down all the equations for the
reactions listed in table I.
\begin{enumerate}
\item The ligand binding reaction is given by:
\begin{equation}
T_{nv}+Ligand
\begin{array}{c}\underrightarrow{k_{f,n}}\\\overleftarrow{k_{b,n}}\end{array} T_{no}
\end{equation}
Since the time scale for ligand binding is much shorter than the
other reactions, the ligand binding reaction can be assumed to be
in quasi-equilibrium and the two populations for each methylation
level can then be written as:
\begin{gather}
  [T_{nv}] = (1-L_n) [T_n], \\
  [T_{no}] = L_n [T_n],
\end{gather}
where $L_n\equiv \frac{[L]}{[L]+K_{d,n}}$, is the receptor
occupancy rate, $[L]$ is the ligand concentration and $[T_n] =
[T_{no}]+[T_{nv}]$ is the total receptor population in methylation
level $n$. The ligand receptor dissociation constant
$K_{d,n}(\equiv k_{b,n}/k_{f,n})$ probably depends on the
methylation level of the receptor $n$ (Dunten, 1991; Borkovich,
1992; Bornhorst, 2000; Sourjik, 2002), however it will become
clear later that this does not affect the perfect adaptation
conditions.

\item The methylation/demethylation reactions can be written as:
\begin{equation}
T_n + E\rightleftharpoons T_nE\rightarrow T_{n\pm 1}+E
\end{equation}
where the enzyme $E$ is either R (CheR) or B (CheB-P). Here we
assume the methylation/demethylation process at the 4 methylation
sites follows a preferred sequence, and therefore the existence of
only 5 methylation states described by $n\in [0,4]$. Though this
assumption is still an open question, it is supported by some
experiments (Shapiro, 1994; Shapiro, 1995). The network of
methylation/demethylation reactions are illustrated in Fig.
\ref{mflow}.

If we assume the above reactions follow Michaelis-Menten kinetics
and the dissociation rates for the bound state are independent of
$\lambda$, i.e., whether the receptor is ligand bound or not, the
bound state concentration can be written as:
\begin{equation}
 [T_nE] =
\frac{[T^F_n][E^F]}{K^E_n}, \label{bound_state}
\end{equation}
where $K^E_n = [(1-L_n){K^E_{nv}}^{-1} + L_n{K^E_{no}}^{-1}]^{-1}$
is the Michaelis constant of the combined (vacant and
ligand-bound) receptor state and the superscript $F$ denotes the
free enzyme and the free substrate (receptor) concentrations.

Since the receptors and the enzymes can exist either in their free
form or bound to each other, the total concentrations of enzymes,
and the concentration of receptors with $n$ methylated sites are
given by the following equations:
\begin{gather}
  [R^T]=[R^F](1+\sn\frac{[T_n^F]}{K^R_n}),\label{Bf1}\\
 [B^{P}]=[B^{PF}](1+\sn\frac{[T_n^F]}{K^B_n}) ,\label{Bf2}\\
 [T_n] = \left(1+\frac{[R^F]}{K^R_n} +
                  \frac{[B^{PF}]}{K^B_n}\right)[T^F_n],
  \label{Bf3}
\end{gather}
where the $[R^T]$, $[B^{P}]$ and $[T_n]$ are the concentrations of
CheR, phosphorylated CheB and receptors with $n$ methyl groups,
respectively.

The kinetic equation for the receptor concentrations $[T_n]$ at
each methylation level can be written as:
\begin{equation}
\frac{d[T_n]}{dt}=J_{n-1}-J_{n}\;\;, \label{dTdt}
\end{equation}
where $J_{n}$ is the net flux from methylation level $n$ to level
$(n+1)$, which is just the difference of methylation and
demethylation rates between these two states. Using the bound
state concentration given in Eq. \ref{bound_state}, $J_n$ can be
written as :
\begin{equation} \label{SS}
  J_{n}=k^R_n\frac{[R^F][T^F_n]}{K^R_n}
     -   k^B_{n+1}\frac{[B^{PF}][T^F_{n+1}]}{K^B_{n+1}},\;\;\;(0\le n\le
     3),
\end{equation}
where $k^R_n$ and $k^B_n$ are the catalytic constants for the
methylation and demethylation reaction respectively, which are
assumed to be independent of $\lambda$, the ligand binding status
of the receptor.  The boundary conditions for the methylation flux
are: $J_{-1}=J_{4}=0$.

\item The auto-phosphorylation of CheA reaction is
given by:
\begin{equation}
  T^{U}\nl \xrightarrow{k^P\nl} T^{P}\nl,
\end{equation}
the phosphate group is subsequently transferred from CheA-P to
CheB and CheY:
\begin{gather}
  T\nl^{P}+Y^{U} \xrightarrow{k^{PY}\nl} T\nl^{U}+Y^P,\\
  T\nl^{P}+B^{UF} \xrightarrow{k^{PB}\nl} T\nl^{U}+B^{PF}.
\end{gather}
While CheB-P dephosphorylates spontaneously, the CheY-P hydrolysis
is enhanced by the phosphatase CheZ, an effect that is included in
the high hydrolysis rate $k^{HY}$ for CheY-P { (Lukat, 1991)}:
\begin{gather}
  Y^P \xrightarrow{k^{HY}} Y,\\
  B^{PF} \xrightarrow{k^{HB}} B^{UF}.
\end{gather}
The kinetic equations for these reactions are:
\begin{gather} \label{YP1}
  \frac{d[Y^P]}{dt}= \sn
  k^{PY}_n[T^{P}_n][Y^U]-k^{HY}[Y^P], \\
   \frac{d[B^{PF}]}{dt}= \sn
  k^{PB}_n[T^{P}_n][B^{UF}]-k^{HB}[B^{PF}],\label{YP2} \\
  \frac{d[T^{P}_n]}{dt} = k^{P}_n [T^U_n]-\nonumber\\
  k^{PY}_n[T^{P}_n][Y^U]-k^{PB}_n[T^{P}_n][B^{UF}]+J_{n-1}^{P}-J_{n}^{P}
  ,
  \label{YP3}
\end{gather}
where $[Y^U]=[Y^T]-[Y^P]$, $[B^{UF}]=[B^{F}]-[B^{PF}]$ and
$[T_n^{U}]=[T_n]-[T_n^{P}]$.  $J_{n}^{P}$ is the net
phosphorylated receptor flux between methylation level n and
$(n+1)$, given similarly as for $J_{n}$ in Eq. \ref{SS} with the
free receptor concentration $[T_n^F]$ replaced by the
phosphorylated free receptor concentration $[T_n^{FP}]$. In all
the above equations, the dependence on $\lambda$ is omitted, so
the autophosphorylation rate and the phosphate transfer rates
should all be considered as the rate for the combined receptor
state (ligand occupied and unoccupied):
$k_n^P=L_nk_{no}^{P}+(1-L_n)k_{nv}^{P}$,
$k_n^{PY}=L_nk_{no}^{PY}+(1-L_n)k_{nv}^{PY}$,
$k_n^{PB}=L_nk_{no}^{PB}+(1-L_n)k_{nv}^{PB}$. It is also assumed
that only CheB-P can bind with the receptors, which leads to the
equation relating different subspecies of CheB:
\begin{equation} \label{BTR}
  [B^T] = [B^{P}] + [B^{F}] - [B^{PF}].
\end{equation}

\end{enumerate}

To describe the kinetics of the signal transduction pathway in
full, we need to consider the interactions among the
concentrations of all the 65 states for the 4 chemical species: 60
receptor states = 2 ligand binding states $\times$ 5 methylation
states $\times$ 3 enzyme binding states $\times$ 2 phosphorylation
states, 1 free CheR state, 2 free CheB states and 2 CheY states.
Using the fact that ligand binding kinetics is fast and the
enzymatic reactions are governed by Michaelis-Menten kinetics, the
number of independent receptor concentrations is reduced from 60
to just 10, consisting of the 5 free methylation states and the 5
phosphorylation states. Now, the whole system is described by
kinetic equations Eq. \ref{dTdt} and Eq. \ref{YP1}-\ref{YP3} plus
conservation equations given by Eq. \ref{Bf1}-\ref{Bf3} and Eq.
\ref{BTR}.

Concentration of the phosphorylated CheY ([Y$^P$]), which
determines the tumbling frequency of bacteria, can be considered
as the output of the whole chemotaxis signal transduction pathway.
In the next section, we study how the steady state concentration
of CheY-P depends on the external ligand concentration $[L]$, in
particular, we {\it derive} a set of conditions for [Y$^P$] to be
independent of $[L]$, i.e., perfect adaptation.

\section*{CONDITIONS FOR PERFECT ADAPTATION}

All the concentrations in our model fall naturally into two
categories: the {\it local} variables defined for one particular
methylation level, such as $[T_n]$, the concentration of receptors
with $n$ methyl groups, and the {\it global} variables, such as
$[R^{F}]$, the concentration of the free CheR.
The system adapts by adjusting the local variables with the ligand
concentration, e. g., the steady-state values of $[T_n]$ varies
with $[L]$. However, perfect adaptation is achieved when the
equilibrium value of [Y$^{P}$], a global variable, is independent
of the ligand concentration { (Othmer, 1998)}. This is
generally not possible because the global variables are coupled
with the local ones. One goal of this paper is to discover the
conditions under which $[Y^P]$ becomes independent of $L$.

The strategy in obtaining the perfect adaptation conditions is to
consider only {\it global} equations, such as the conservation
equations of the chemical species (e.g., Eq. \ref{Bf1}, \ref{Bf2}
and \ref{BTR}) and the steady-state equations of global variables
(e.g., Eq. \ref{YP1} and \ref{YP2}), which do not depend on any
one specific methylation level. In these global equations, there
is no explicit dependence on ligand concentration, and {\it
composite} variables, such as $\sn \frac{[T_n^F]}{K_n^{R}}$ in Eq.
\ref{Bf1}, enter as weighted sums of the methylation level
specific receptor concentrations.
Another kind of global equation can be constructed by summing
steady-state equations at all methylation levels (e.g., Eq.
\ref{Bf3}, \ref{SS} and \ref{YP3}). The price to pay for such
global equations is the introduction of new composite variables.
However, if the reaction rates involved in different reactions are
related in certain ways, the {\it same} composite variables appear
in different global equations so that there are enough global
equations to determine all the independent global and composite
variables. In other words, if certain conditions between reaction
rates are satisfied, the steady-state concentrations of all the
global and composite variables including $[Y^P]$ can be
independent of the ligand concentration, i.e., perfect adaptation.

We leave the detailed derivation for the perfect adaptation
conditions to the appendix. In the following, we list these
conditions, discuss their meaning and compare them with those
found in previous works (Barkai, 1997; Yi, 2000). The perfect
adaptation conditions can be grouped for each of the three pathway
processes: condition \ref{Cond:LFast} is for the ligand binding
and unbinding, conditions \ref{Cond:KLinear}-\ref{Cond:rkRkB} are
required for the methylation process and conditions
\ref{Cond:BYP2}-\ref{Cond:rRB} are related to the phosphorylation
process:

\begin{enumerate}

\item\label{Cond:LFast} The time scale for ligand binding
is much shorter than the methylation and phosphorylation time
scale. This condition allows us to neglect ligand
binding/unbinding kinetics.

\item\label{Cond:KLinear} The association
rates between the receptor and the methylation/demethylation
enzymes, CheR and CheB-$P$, are linearly related to the activity
of the receptor and are zero for $n=4$ and $n=0$, respectively:
${K_{\nl}^{R}}^{-1}\propto P_{4\lambda}-P\nl$ and
${K_{\nl}^{B}}^{-1}\propto P\nl-P_{0\lambda}$. The dissociation
rates of the enzyme receptor bound states are independent of
$\lambda$.

\item \label{Cond:CloseLoop}The
receptor activities of the non-methylated and the maximally
methylated receptors are independent of $\lambda$:
$P_{0v}=P_{0o}$, $P_{4v}=P_{4o}$.

\item \label{Cond:rkRkB} The ratios between the CheR catalytic rate ($k^R_n$)
and the CheB-P catalytic rate of the next methylation level
($k^B_{n+1}$) are the same for all methylation states $n$:
$k^B_{n+1}/k^R_{n}=\text{const.}$.

\item \label{Cond:BYP2} The phosphate transfer rates from CheA to CheB or
CheY are proportional to CheA auto-phosphorylation rate:
$k_{\nl}^{PB}\propto P_{\nl}$, $k_{\nl}^{PY}\propto P_{\nl}$.

\item \label{Cond:rRB}The explicit dependence on [$T_n^F$] distribution can
be removed from the expression
\begin{equation}  \label{RBR}
  \xi \equiv \left(-\frac{[R^{F}]}{K^R}+\frac{[B^{PF}]}{K^B}\right) \sn
  {P_n}^2 [T^F_n].
\end{equation}
This condition can only be strictly satisfied when $
\frac{[R^{F}]}{K^R}=\frac{[B^{PF}]}{K^B}$.
\end{enumerate}

Condition \ref{Cond:LFast} is necessary to decouple the ligand
binding process from the rest of the reactions. This is verified
experimentally and assumed in all the previous models (Barkai,
1997; Morton-Firth, 1998; Spiro, 1997; Yi, 2000).

Condition 2 for the methylation process requires that the CheR and
CheB methylation/demethylation rates depend linearly on the
receptor's auto-phosphorylation rate (activity) . This is a
generalization of the key ingredient for perfect adaptation found
in Barkai and Leibler's work (Barkai, 1997). In the special case
of $P_{4\lambda}=1$ and $P_{0\lambda}=0$, condition
\ref{Cond:KLinear} means that CheB-P only bind to active receptors
and CheR only bind to inactive receptors, the latter is missed in
the original work of BL and later found to be necessary for
perfect adaptation in (Morton-Firth, 1999) through a direct
numerical simulation of the full system.

The requirement in condition 3 that $P_{0\lambda}$ and
$P_{4\lambda}$ be independent of $\lambda$ is needed so that both
the ligand-bound and vacant receptors have the same range of
activity. This requirement for perfect adaptation is necessary in
case the extreme methylation states $n=0$ or $n=4$ become
populated with receptors.

Condition 4 was first pointed out in (Yi, 2000), it is a more
general form of the assumption that both $k_{n}^{R}$ and
$k_{n}^{B}$ are independent of $n$ made in the original BL model.
The justification of this condition may be related to a common
evolutionary origin of CheR and CheB, resulting in a similar
anchoring position to the receptor for CheR methylating site $n$
and CheB-P demethylating site $n+1$ (Shapiro, 1994; Shapiro, 1995;
Djordjevic, 1998;  Barnakov, 1999).

Condition 5 for the phosphorylation process is very similar to
condition 2, in the sense that the phosphate transfer rates of the
receptors have to be linearly related to their activity. This
condition was not discovered before because the phosphorylation
process was neglected in previous works (Barkai, 1997; Yi, 2000).

Condition 6 can only be satisfied exactly when one tunes the
parameters such that the pre-factor in front of the sum in Eq.
\ref{RBR} is zero. This condition was overlooked by most of the
previous studies because the activities of the CheR or CheB-P
bound receptors were neglected. However, in equilibrium, the
population of enzyme bound receptors can be as high as $30\%$
(Morton-Firth, 1999).

\begin{table*}
\caption{System parameters, numerical values from (Morton-Firth, 1999).}
\label{Tab:rates}
\begin{center}

\begin{tabular}{llr}
\hline
Symbol & Description & Value\\
\hline $P\nl$ & Relative activity of $T\nl$ &
\begin{tabular}{l|lllll}
&0&1&2&3&4\\ \hline $v$&0&0.125&0.5&0.874&1\\ $o$&0&0.017&0.125&0.5&1\\
\end{tabular} \\
$K^R$  & CheR Michaelis constant & $0.364\,\mu\text{M}$\\
$K^B$  & CheB Michaelis constant & $1.405\,\mu\text{M}$\\
$k^R$  & CheR catalytic constant & $0.819\,\text{s}^{-1}$\\
$k^B$  & CheB catalytic constant & $0.155\,\text{s}^{-1}$\\
$k^P$  & CheA autophosphorylation rate & $15.5\,\text{s}^{-1}$\\
$k^{PY}$ & CheA$\rightarrow$CheY phosphorus transfer rate & $5\,\mu\text{M}^{-1}\text{s}^{-1}$\\
$k^{PB}$ & CheA$\rightarrow$CheB phosphorus transfer rate & $5\,\mu\text{M}^{-1}\text{s}^{-1}$\\
$k^{HY}$ & CheY dephosphorylation rate & $14.15\,\text{s}^{-1}$\\
$k^{HB}$ & CheB dephosphorylation rate & $0.35\,\text{s}^{-1}$\\
\hline
\end{tabular}
\end{center}
\end{table*}

By imposing all the conditions above, the steady state
concentrations of the global variables will be independent of the
ligand concentration, and are determined by 15 parameters: the 4
total concentrations of table \ref{Tab:species}, and 11 reaction
rates of table \ref{Tab:rates}, including $P_{4}$ and $P_0$, but
not the relative activity values for the rest of the methylation
states. However, for real biological system, these conditions for
perfect adaptation may not be strictly satisfied. In order to
understand bacteria's ability in adapting accurately under
different internal and external conditions, i. e., robustness, we
need to evaluate the effect of violating these perfect adaptation
conditions.

\section*{EFFECTS OF VIOLATING THE PERFECT ADAPTATION CONDITIONS}

\begin{figure}
\includegraphics{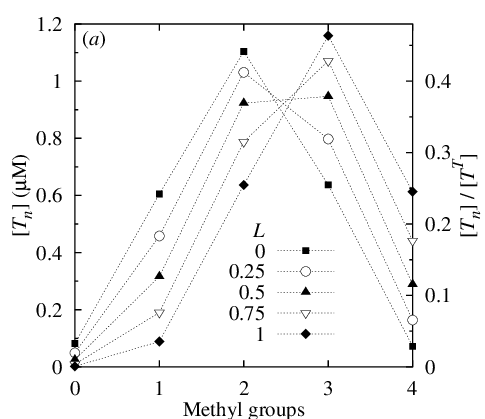}\\
\includegraphics{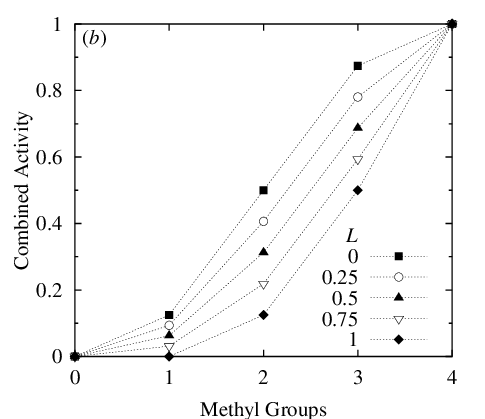}
\caption{($A$) Distribution of receptors in different methylation
states at different ligand occupancy fractions $L$ for the
reference parameters. The total activity of the system being
$[T^{A}]=0.5[T^T]$. ($B$) The population-weighted average receptor
activity $P_n(L)$ for different methylation level $n\in[0,4]$ at
different fractional ligand occupancy rates $L$.}
\label{Fig:tdcondBYP}
\end{figure}

\begin{table}[h]
\label{Table:con} \caption{Protein concentrations (in $\mu$M) at
different ligand occupancy rates $L$ for the reference
parameters.}
\begin{center}

\begin{tabular}{crrr}
 \hline
 Species  & $L=0$ & $L=0.5$ & $L=1$ \\
 \hline
 $[T_0]$   & 0.028 & 0.025 & 0.002 \\
 $[T_1]$   & 0.605 & 0.316 & 0.089 \\
 $[T_2]$   & 1.104 & 0.923 & 0.637 \\
 $[T_3]$   & 0.637 & 0.947 & 1.159 \\
 $[T_4]$   & 0.072 & 0.289 & 0.613 \\
 $[T^A]$   & 1.257 & 1.250 & 1.274 \\
 $[T^P]$   & 0.202 & 0.201 & 0.204 \\
 $[R^F]$   & 0.050 & 0.050 & 0.050 \\
 $[B^F]$   & 1.603 & 1.602 & 1.603 \\
 $[B^{PT}]$& 1.858 & 1.857 & 1.860 \\
 $[B^{PF}]$& 1.191 & 1.190 & 1.193 \\
 $[Y^P]$   & 1.200 & 1.196 & 1.209\\
 \hline
\end{tabular}
\end{center}
\end{table}

Since it is not feasible to explore the whole parameter space, we
choose to mostly perturb around the parameter values that have
been used in previous studies. To this end, we take most of our
parameters from (Morton-Firth, 1999), which are listed here in
Table II and Table III. Hereafter we refer to this set of
parameters as the reference parameters. Assuming ligand occupancy
rate $L_n=L$ is independent of $n$, the steady state receptor
distributions in different methylation states for different ligand
occupancy rates $L$ is shown in Fig. \ref{Fig:tdcondBYP} $A$ for
the reference parameters. In Fig. \ref{Fig:tdcondBYP} $B$, the
population-weighted average receptor activities
$P_n(L)=P_{no}L+P_{nv}(1-L)$ for methylation level $n\in [0,4]$ is
also shown. As is clear from Fig. \ref{Fig:tdcondBYP}, when ligand
(attractant) occupancy rate increases, the average receptor
activity $P_n(L)$ decreases for each methylation level $n$, and
the system adapts by shifting the receptor population towards
higher methylation states in achieving constant total activity
$[T^A]=\sn P_{n}(L)[T_n]$. The steady state concentrations of all
the other relevant concentrations at 3 different ligand occupancy
fractions are given in Table IV for the reference parameters, the
small changes in $[Y^P]$ at different ligand concentrations are
caused by violation of conditions 5 and 6 in the reference model
used in (Morton-Firth, 1999) as we explain later in the section
``Violating condition 5".

{ We have also constructed another model by modifying some of
the reference parameters so that all the perfect conditions are
satisfied. The results of perturbing this new model are
essentially the same as for the reference model, mainly because
the adaptation error in the reference model is very small
($<1\%$). While this new model is mathematically more rigorous for
isolating different error sources, the reference model has the
advantage that it is motivated biologically (from experiments or
common sense), and therefore serves as a better starting point in
exploring the parameter regions that are more likely to be
biologically relevant. To make sure violation of conditions 5 and
6 in the reference model does not contaminate the effect of other
conditions too much, we have always checked the error with and
without violating the condition in consideration, and made sure
most of the error does come from violating the perfect
condition we study.} 

Since ligand binding is much faster than other relevant processes
of the system, we do not consider the unrealistic situation of
violating condition 1. In the following, we study the effects of
breaking the other 5 perfect adaptation conditions. { Our goal
is to understand the general reason behind the robustness of the
system with respect to breaking each perfect adaptation condition.
Even though we primarily perturb the system around the reference
parameters, we also explore other parameter regions, especially
when the reference model becomes insensitive to violation of a
given condition. This strategy allows us to gain the general
understanding of where in the parameter space a given perfect
adaptation condition becomes important and the reason behind it.}

\subsection*{Violation of condition 2}

Condition 2 requires that the methylation/demethylation enzyme
binding rates to a receptor depend linearly on the activity of the
receptor. For the reference parameters, where $P_0=0$ and $P_4=1$,
condition 2 simply means that CheR only binds to inactive
receptors and CheB-P only binds to active receptors. The simplest
way in violating condition 2 is to allow CheR bind to active
receptor or CheB-P bind to inactive ones , which can be formally
expressed as:
\begin{gather}
 {K_{\nl}^R}^{-1}={K^R}^{-1}b_r(1-P_{\nl}+a_r),\nonumber
\\
{K_{\nl}^B}^{-1}={K^B}^{-1}b_b(P_{\nl}+a_b), \label{cond2}
\end{gather}
where $a_{r}\ge 0$ and $a_{b}\ge 0$ are the measures of violating
condition 2; $b_{r}$ and $b_{b}$ are normalization factors tuned
with respect to $a_{r}$ and $a_b$ to keep the total activity of
the system constant at a given ligand occupancy rate (L=0.5) for
comparison purpose. $a_r=0$ and $a_b=0$ corresponds to condition 2
being satisfied; $a_r\rightarrow\infty$ (with $a_rb_r=const.$) or
$a_b\rightarrow\infty$ (with $a_bb_b=const.$) respectively
corresponds to CheR or CheB-P binding to all receptors equally.

\begin{figure}
\includegraphics{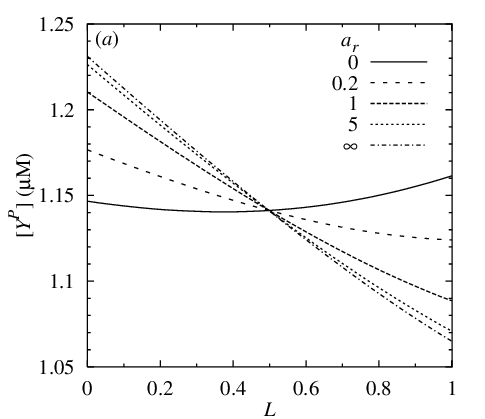}
\includegraphics{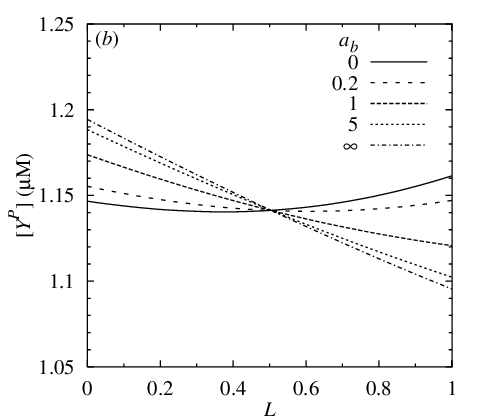}
\caption{The steady state $[Y^P]$ concentration versus ligand
binding rate $L$ for different ways of breaking condition 2: ($A$)
CheB-P binds with active receptors only ($a_b=0$), CheR is allowed
to bind with active receptor with varying strength
$a_r=0,.2,1,5,\infty$, where $a_r=\infty$ corresponds to CheR
binds to all receptor indiscriminately. ($B$) CheR binds with
inactive receptors only ($a_r=0$), CheB-P is allowed to bind with
inactive receptor with varying strength $a_b=0,.2,1,5,\infty$,
where $a_b=\infty$ corresponds to CheB-P binds to all receptor
indiscriminately.} \label{Fig:Cond2}
\end{figure}

In Fig. \ref{Fig:Cond2}, we show the steady-state concentration
of CheY-P versus the ligand occupancy rate $L$ for various values
of $a_r$ and $a_b$. Even for the extreme cases of $a_{r}=\infty$
or $a_{b}=\infty$, respectively corresponding to CheR or CheB-P
binding to both active and inactive receptors equally, the
deviation from perfect adaptation is only $\sim 10\%-15\%$.
Intuitively, the reason for the near perfect adaptation is that
the control of the system's total activity can be carried out by
either the methylation (CheR) or demethylation (CheB-P) process,
provided that at least one of the enzymes' binding rates is
strongly correlated with the receptor activity. If the receptor
binding rates of both enzymes become independent of the receptor's
activity, i. e., both $a_{r}=\infty$ and $a_{b}=\infty$, the
system is only controlled through the weak effect of CheB
phosphorylation and does not adapt very well.

Specifcally, condition 2 requires that CheR does not bind to the
fully methylated receptors (n=4), and CheB-P does not bind to the
unmethylated receptors (n=0). Therefore, the quantitative effects
of breaking condition 2 (as in Eq. \ref{cond2}) depends on the
receptor concentration at the fully methylated state $[T_4]$ or
the unmethylated states $[T_0]$ (see appendix for details). Both
$[T_0]$ and $[T_4]$ are relatively small for the reference
parameters with $[T_4]>[T_0]$ (see Fig. \ref{Fig:tdcondBYP}),
which explains the qualitative features in Fig. \ref{Fig:Cond2}.
The effect of $a_b\rightarrow \infty$ only becomes noticeable
because $[T_0]$ is not too small for $a_b\rightarrow \infty$.

\subsection*{Violation of condition 3}

\begin{figure}
\includegraphics{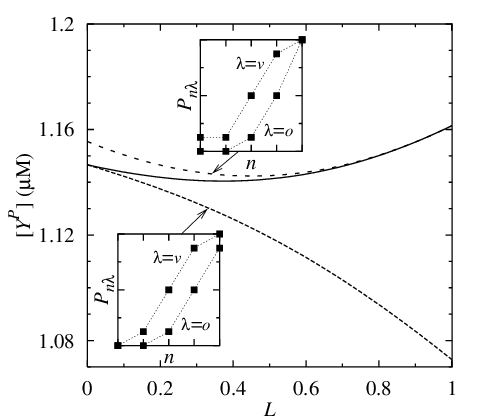}
\caption{The steady state $[Y^P]$ concentration versus ligand
occupancy fraction $L$ for different ways of breaking condition 3
by opening the activity gap at $n=0$: $P_{0v}=P_{1v}=\frac{1}{8}$
(long dashed line) or at $n=4$: $P_{4o}=P_{3v}=\frac{7}{8}$ (short
dashed line), the solid line is for the reference parameters. The
two inserts illustrate the opening of the activity gap at $n=0$
and $n=4$ respectively. } \label{Fig:Cond3}
\end{figure}

Since adaptation for bacterial chemotaxis relies on balancing the
effect of ligand binding on the receptor's activity with that of
the methylation of the receptor, a necessary condition for perfect
adaptation is for both ligand bound and vacant receptors to have
the same range of activity, i.e., condition 3. For the reference
parameters, condition 3 is obeyed by having: $P_{0v}=P_{0o}=0$;
$P_{4v}=P_{4o}=1$. Without changing the monotonic dependence of
the receptor activity on their methylation level, we can break
condition 3 at $n=0$ by increasing $P_{0v}$ from $0$ to
$\frac{1}{8}$; or at $n=4$ by decreasing $P_{4o}$ from $1$ to
$\frac{7}{8}$. The enzyme binding rates are adjusted accordingly
in keeping condition 2 satisfied. The effects are shown in Fig.
\ref{Fig:Cond3}. The system is insensitive to the opening of the
activity gap $\Delta P_0\equiv P_{0v}-P_{0o}$ at $n=0$, because
the receptor population is small at $n=0$ even at $L=0$. For the
same opening of activity gap $\Delta P_4\equiv P_{4v}-P_{4o}$ at
$n=4$, the adaptation error is $6\%$. In particular, the system
has lower CheY-P concentration at higher ligand occupancy rate
$L$, because the receptor population shifts towards higher
methylation levels at larger $L$, and the effect of methylation is
not large enough to cancel the decrease of activity caused by
ligand binding. Quantitatively, the adaptation error increases
with the activity gap; e.g., it reaches $25\%$ when we lower
$P_{4o}$ further to $0.5$.

\subsection*{Violation of condition 4}

\begin{figure}
\includegraphics{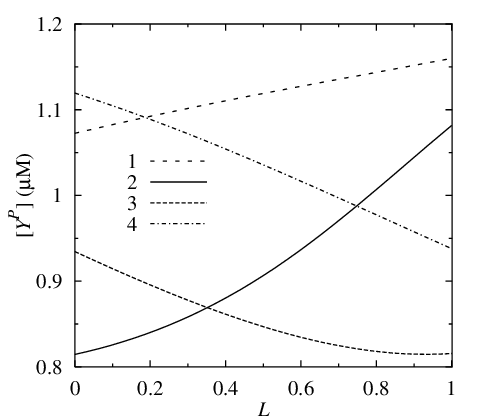}
\caption{Steady-state CheY-P concentration $[Y^P]$ versus ligand
occupancy rate $L$ for breaking condition 4. { The results are
obtained by increasing one of the $4$ ratios of catalytic rates
(see text for definition) by a factor of $2$: $r_n=2r_{ref}=0.38$,
while keeping the other $3$ ratios unchanged at the reference
value of $0.19$. The $4$ curves correspond to $n=1,2,3,4$
respectively.}} \label{Fig:Cond4}
\end{figure}

The methylation and demethylation catalytic rates $k_{n}^R$ and
$k_{n}^B$ can depend on methylation level $n$. From Eq. \ref{SS},
the steady-state properties of the system only depend on the
ratios: $r_n=k_{n}^B/k_{n-1}^R$ for $n=1,2,3,4$. Condition 4 for
perfect adaptation requires that $r_n$ be a constant independent
of $n$, a kind of ``detailed balance" condition. Indeed, if we
change $k_{n}^R$ and $k_{n}^B$ while keeping $r_n$ constant, the
system adapts perfectly. However, when we make $r_n$ depend on
$n$, perfect adaptation is lost. In Fig. \ref{Fig:Cond4}, we show
the effects of increasing one $r_n$ by a factor of 2 while keeping
the rest $r_n$ unchanged at their reference value for $n=1,2,3,4$
respectively. The quantitative deviation from perfect adaptation
depends on $n$, with the largest deviation of $\sim 25\%$
occurring at $n=2$ , possibly because the receptors are highly
populated at $n=2$ for the reference parameters.

\subsection*{Violating condition \ref{Cond:BYP2}}

Condition 5 requires that the phosphate transfer rates of a
receptor is proportional to its auto-phosphorylation rate, a kind
of compatibility condition. The simplest way to break condition
\ref{Cond:BYP2} is to set the phosphate transfer rates to be a
constant independent of both the ligand binding and the
methylation level of the receptor. This assumption is also made in
(Morton-Firth, 1998; Morton-Firth, 1999).

\begin{figure}
\includegraphics{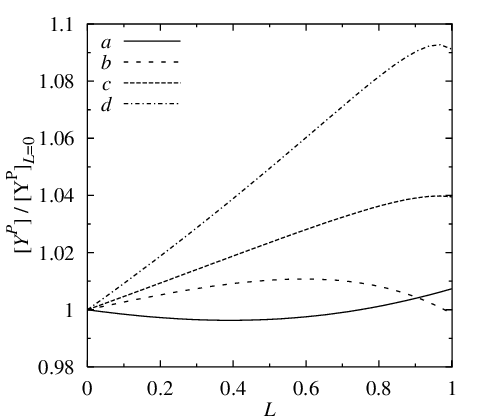}
\caption{Relative steady-state CheY-P concentrations
$[Y^P]/[Y^P]_{L=0}$ versus ligand occupancy rate $L$ when
condition \ref{Cond:BYP2} is violated; the adaptation error
depends on the parameters of the system, 4 cases are studied here
using parameters with increasing degrees of deviation from their
reference values (see text for detail). The parameters used are:
(1) For curve $a$, reference parameter values; (2) For curve $b$,
same as for curve $a$, except $[Y^T]=[Y^T]\re/20$; (3) For curve
$c$, same as for curve $b$, except $k^R=k^R\re/50$; (4) For curve
$d$, same as for curve $c$, except $P_{v1}=0.25$ and
$P_{v2}=0.6$.} \label{Fig:condBYP}
\end{figure}

For the reference parameters, the steady-state $[Y^P]$ change by
less than 1\% over the whole range of ligand occupancy as shown in
Fig. \ref{Fig:condBYP} (curve $a$), indicating the insensitivity
of the system's perfect adaptation with respect to this particular
choice of breaking condition \ref{Cond:BYP2}. In the following, we
explain the system's near perfect adaptation by the existence of
approximate global equations.

In deriving condition \ref{Cond:BYP2}, a global equation is formed
by summing Eq. \ref{YP3} over all methylation levels and replacing
$[T_n^U]$ by $[T_n]-[T_n^{P}]$, which leads to the formation of 4
composite variables: $G_0=\sn k^{P}_n [T_n]$, $G_1=\sn
k^{P}_n[T_n^{P}]$, $G_2=\sn k^{PY}_n[T_n^{P}]$ and $G_3=\sn
k^{PB}_n[T_n^{P}]$. Condition \ref{Cond:BYP2} is needed to make
$G_1$, $G_2$ and $G_3$ proportional to each other, so that the
total number of global equations is enough to solve for all the
independent global and composite variables (see section
``Conditions for perfect
adaptations" 
and Appendix for details). When condition \ref{Cond:BYP2} is
broken by setting $k^{PB}_n$ and $k^{PY}_n$ to be constant, $G_2$
and $G_3$ are still proportional to each other, but they are now
different from $G_1$, the total number of global equations are now
not enough in solving for all the global variables, and local
equations have to be used. This leads to all the global variable
depend on ligand concentration, i.e., non-perfect adaptation.
However, because the concentration of (un-phosphorylated) CheY is
much larger than the receptor concentrations, the phosphorylated
receptor concentration $[T^{P}_n]$ is small compared with the
total receptor concentration $[T_n]$, due to efficient phosphate
transfer from CheA to CheY and the subsequent high CheY-P
dephosphorylation rate. As a result, $G_1$ is negligible relative
to $G_0$, leading to an approximate global equation with the same
degree of reduction in independent composite variables and
eventually the near perfect adaption observed in Fig.
\ref{Fig:condBYP} $A$.

However, reducing CheY concentration alone does not change too
much the system's ability in perfect adaptation, as shown in Fig.
\ref{Fig:condBYP} (curve $b$). At low CheY concentration, the
phosphate group of CheA-P goes to CheB. Because of the slow
dephosphorylation rate of CheB-P, most of the CheB become
phosphorylated in steady state, essentially decoupling the
phosphorylation process from the adaptation process. The
adaptation of the system therefore becomes insensitive to the
phosphorylation related condition 5.

To amplify the effect of violating condition $5$, we reduce the
overall activity to $[T^A]=0.014[T^T]$ at $L=0$ by making
$k^R=0.02k^R\re$. The result is shown in Fig. \ref{Fig:condBYP}
(curve $c$). The adaptation accuracy can also depend on other
parameters, such as the receptor activity $P\nl$. In Fig.
\ref{Fig:condBYP} (curve $d$), we show that a slight change in
receptor activity leads to higher deviation from perfect
adaptation.

\subsection*{Violating condition \ref{Cond:rRB}}

The total receptor activity $[T^{A}](\equiv\sn P_n [T_n])$ is
directly related to the final production of CheY-P. However, only
part of $[T^{A}]$ can be expressed in terms of other composite
variables related to receptor population, i. e., the total free
receptor concentration $[T^{F}]\equiv\sn [T^F_n]$ and the total
activity due to free receptors $[T^{AF}]\equiv\sn P_{n}[T^F_n]$.
It has an extra term $\xi$ coming from the activity of the enzyme
(CheR or CheB-P) bound receptors (see appendix for details), which
is proportional to $\xi'=\sn P^2_{n}[T^F_n]$ with a pre-factor
$\left( -\frac{[R^{F}]}{K^R}+\frac{[B^{PF}]}{K^B} \right)$ (see
Eq. \ref{RBR}). Condition \ref{Cond:rRB} is required to eliminate
this extra global variable $\xi'$ by setting the pre-factor to
zero.

\begin{figure}
\begin{center}
\includegraphics[width=6.6cm]{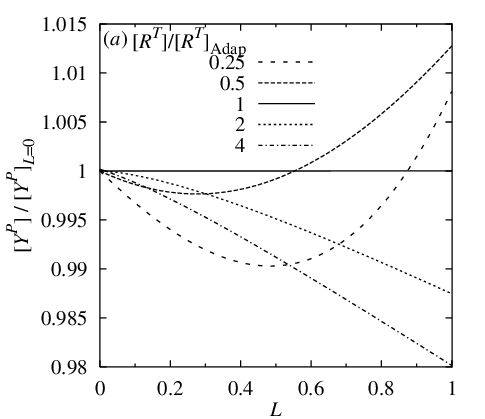}
\includegraphics[width=6.6cm]{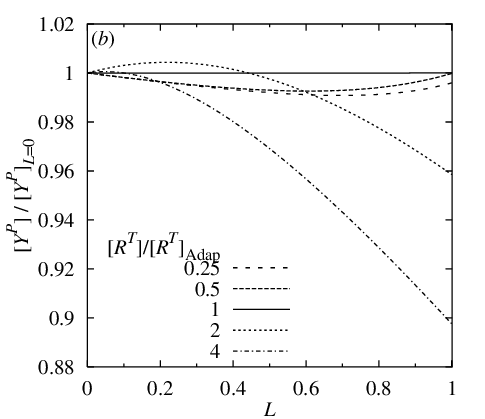}
\includegraphics[width=6.6cm]{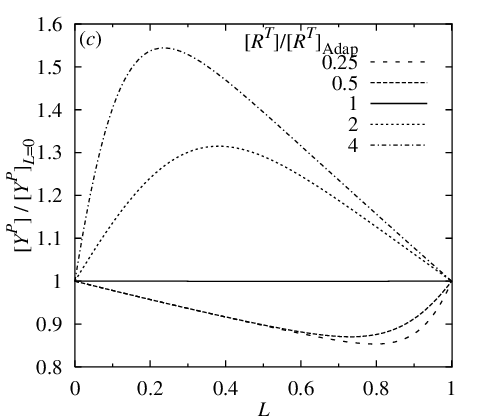}
\end{center}
\caption{Relative steady-state CheY-P concentrations
$[Y^P]/[Y^P]_{L=0}$ versus ligand occupancy rate for different
CheR concentrations (with condition 5 satisfied), which are varied
with respect to the perfect adaptation value $[R^T]_{Adap}$: ($A$)
Reference parameters are used except the different values of
$[R^T]$ listed in the figure, and
$[R^{T}]_{Adap}=2.63[R^T]_{ref}$; ($B$) $k^{R}=0.1k^{R}_{ref}$ is
chosen in reducing the total activity, where adaptation is less
accurate, and $[R^{T}]_{Adap}=5.26[R^T]_{ref}$; ($C$) Same
parameters as in $B$ except that the activity difference between
ligand bound and vacant receptors are set to be maximum (see
text), and $[R^{T}]_{Adap}=5.35[R^T]_{ref}$.} \label{Fig:condRB0}
\end{figure}

The effect of breaking condition \ref{Cond:rRB} can be small,
because as $[R^F]$ deviates from its perfect adaption value
$[R^F]_{Adap}$, so does $[B^{PF}]$ with the same trend, leading to
small changes of the pre-factor in $\xi$. Also, part of $\xi'$ can
be approximated by a linear combination of $[T^{F}]$ and
$[T^{AF}]$, depending on the activity levels of different
receptors $P\nl$. Finally, for higher total activity, the relative
effect of $\xi$ will be small. For the reference parameters, the
accuracy of adaptation is better than $98\%$ for 4-fold change of
CheR concentration from its perfect adaptation value, as shown in
Fig. \ref{Fig:condRB0} $A$. The adaptation accuracy decreases as we
lower the total activity by decreasing methylation rate $k^{R}$,
as shown in Fig. \ref{Fig:condRB0} $B$. Finally, when we increase
the activity differences between the ligand bound and the vacant
receptors by setting: $P_{no}=0$ $(n=0,1,2,3)$, $P_{4o}=1$;
$P_{nv}=1$ $(n=1,2,3,4)$, $P_{0v}=0$, the same change in $[R^T]$
can cause more than a 50\% error in adaptation, as shown in Fig.
\ref{Fig:condRB0} $C$.

\section*{COMPARISON TO STOCHASTIC SIMULATION AND EXPERIMENTS}

The results from the previous sections can be compared with both
the discrete stochastic numerical simulation and real experiments.
We use the reference parameters for all the comparison studies.

\subsection*{Comparison to stochastic simulation }

Stochsim (Morton-Firth, 1998) is a general purpose stochastic
simulator for chemical reactions. For our study, the volume of
Stochsim simulation is set to be $1.4\time 10^{-15}\,\text{L}$,
and the number of molecules is therefore $843\times$concentration
(in $\mu$M).

\begin{figure}
\includegraphics{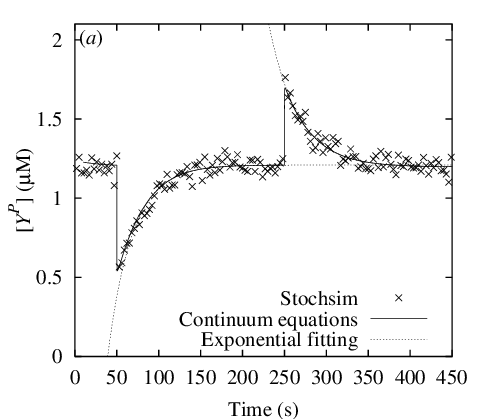}
\includegraphics{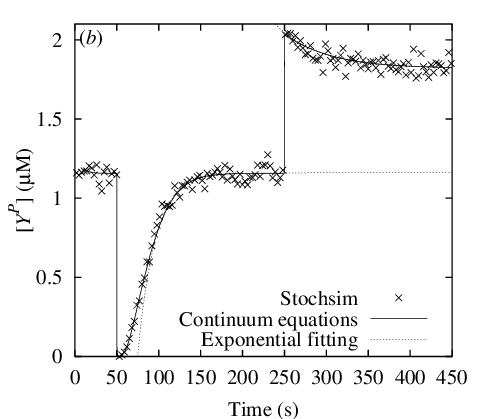}
\caption{Dynamics of [Y$^P$] from Stochsim simulation with ligand
occupancy rates $L$ changing from $0\rightarrow 1\rightarrow 0.2$
at 50 and 250 seconds when the parameters are set to: ($A$) the
reference values, ($B$) same as in Fig. \ref{Fig:condRB0} $C$ with
$[R^T]=4[R^T]_{Adap}$. The solid lines are results from
simulations of our deterministic equations, the dotted lines are
fits to the Stochsim data with an exponential decaying function to
obtain the relaxation time. } \label{Fig:snadap}
\end{figure}

In Fig. \ref{Fig:snadap} $A$, we show the Stochsim simulation
result for the reference parameters, which agrees well with the
results from simulating our continuum equations with the same
parameters. In Fig. \ref{Fig:snadap} $B$, we show the Stochsim
simulation result for the parameters used in Fig. \ref{Fig:condRB0} $C$ with $[R^T]=4[R^T]_{Adap}$, where perfect
adaptation is {\it lost} because of violation of condition 6. As
predicted from our deterministic model, after sudden changes of
ligand occupancy rate $L$, $[Y^{P}]$ does not always return to its
pre-stimulus level; in fact, the maximum error ($\sim50\%$) is
observed when $L=0.2$, consistent with Fig. \ref{Fig:condRB0} $C$.

For most of the results shown in this paper, we have compared with
the results from stochastic simulation using Stochsim (data not
shown). Overall, the averaged behaviors of Stochsim simulations
are consistent with our continuum model, which is interesting
given the nonlinear nature of the chemical kinetics. Further work
is needed in characterizing the fluctuation of the individual
Stochsim simulations, and compare them with the fluctuations in
behavior among different individual bacteria (Morton-Firth, 1998).

\subsection*{Comparison with experiment}

In a recent experimental study by Alon {\it et al} (Alon, 1999),
mutant bacteria lacking a certain chemotaxis protein, such as
CheR, CheB, CheY or CheZ, are used, and the missing protein is
reintroduced in a controlled fashion through a plasmid inserted
into the mutant bacteria cells. This technique allowed these
authors to study the effect of various enzyme concentration
changes on the chemotaxis behavior of the bacteria. Specifically,
the tumbling frequency of the bacteria is measured through a
sudden increase of ligand concentration, which effectively
corresponds to a sudden change of ligand occupancy rate from $L=0$
to $L=1$.

\begin{figure}
\includegraphics{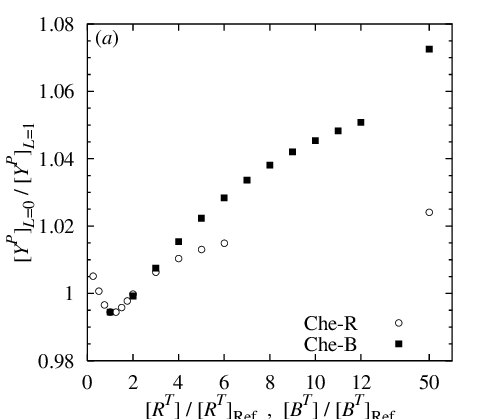}
\includegraphics{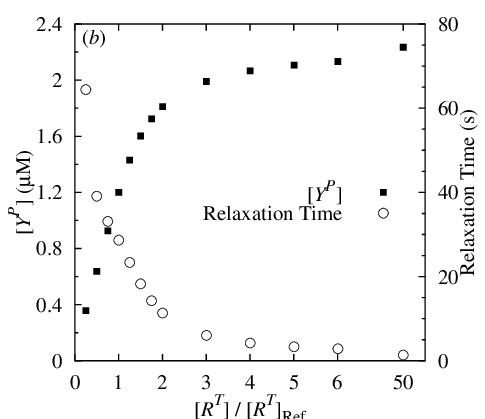}
\caption{The response to a sudden increase of ligand concentration
determined from the continuum model. ($A$) The steady-state CheY-P
concentration ratios before and after the stimulus
$[Y^P]_{L=0}/[Y^P]_{L=1}$ for different fold changes of CheR or
CheB concentrations; ($B$) Steady state CheY-P concentration and the
linear relaxation time upon sudden change of ligand occupancy rate
(from 0 to 1) versus different CheR concentrations.}
\label{Fig:adapsteady}
\end{figure}

In Fig. \ref{Fig:adapsteady} $A$, we show the adaptation
precision as the ratio between phosphorylated CheY level before
and after the stimulus for various CheR and CheB concentrations.
For CheR concentration change of up to 50 fold with respect to the
reference value, the adaptation error is $\le 3\%$, somewhat
smaller than the experimentally measured adaptation error cited in
(Alon, 1999). If $[B^T]$, instead of $[R^T]$, is changed, the
adaptation error would be much bigger, as shown in Fig.
\ref{Fig:adapsteady} $A$. This is the case because for large values
of $[B^T]$, the low activity and the large values of $[B^{PF}]$
make the violation of condition 5 and 6 more significant. This
could  explain the larger (1.09) adaptation precision reported in
(Alon, 1999) when $[B^T]$ expression is 12 times that of the wild
type values. Since we define adaptation accuracy based on CheY-P
concentration, the quantitative difference between the adaptation
error observed in (Alon, 1999) and those of our model could be
explained by the signal amplification at the motor level (Cluzel,
2000).

The relaxation time of the system after a sudden change in ligand
concentration can be determined by direct simulation of the full
kinetic equation or by linearizing the methylation/demethylation
kinetic equations around the steady state. The dependence of both
the steady-state tumbling frequency and the linear relaxation time
on CheR concentration $[R^T]$ is shown in Fig.
\ref{Fig:adapsteady} $B$. They agree qualitatively with the steady
state tumbling frequency and the relaxation time measured in
(Alon, 1999), as depicted in Fig. 2 $B$ of their paper, although
direct quantitative comparison is not possible due to different
definitions of relaxation time and lack of detailed understanding
on how CheY-P regulates the motor.

\section*{DISCUSSIONS AND CONCLUSIONS}

In this paper, we have studied a theoretical model describing the
full chemotaxis signal transduction pathway. Through systematic
analysis of the steady state properties of the model, we {\it
derive} a complete set of conditions for the system to adapt
exactly. Some of the conditions are generalizations of the ones
discovered before, but others, in particular, the conditions
related to the phosphorylation part of the pathway, are discovered
for the first time here. It is quite remarkable that perfect
adaptation can be achieved for arbitrary ligand concentration with
a small set of conditions, far less than the number of variables
and the number of reaction rate constants in the problem.

The (intrinsic) state of a receptor can be described by its ligand
binding status ($\lambda$) and methylation level ($n$). The
(external) properties of the receptor complex include its
abilities to interact with the methylation/demethylation enzymes,
to undergo autophosphorylation, and to transfer its own phosphate
group to CheY or CheB, all of which depends on the (internal)
state of the receptor characterized by $n$ and $\lambda$. Perfect
adaptation requires these three properties of the receptor complex
to be correlated with each other in a linear fashion for any given
receptor state $\{n\lambda\}$ (condition 2 and 5). Available
experimental data that addresses the validity of such connections
has been discussed extensively in (Yi, 2000). Even though the
evidence for such connections is not well established and the
correlation may not be linear, it is conceivable that a high
degree of correlation exists among these three properties of the
receptor, because they are determined by the same conformational
change of the receptor protein complex for a given receptor state
$\{n\lambda\}$.

Since most of the perfect adaptation conditions are relations
between different reaction rates, the system's ability to adapt
accurately can be considered ``robust" in the sense that the
perfect adaptation is independent of concentrations of any
specific chemotaxis protein, which can fluctuate between different
individual cells and at different stage of the cell development.
Only one of the perfect adaptation conditions requires the fine
tuning of the methylation enzyme concentrations (condition 6).
Because of this condition, in the strict mathematical sense, the
perfect adaptation of the system can only be achieved via fine
tuning of a parameter, and therefore cannot be considered robust.
However, as we have shown in this paper, the effect of violating
this condition can be rather small, especially at the reference
parameters.

The discovery of the perfect adaptation conditions provides an
invaluable starting point in exploring the parameter space. We
evaluate the sensitivity of the system's perfect adaptation
ability by perturbing the perfect adaptation conditions. We find
that the system can adapt near perfectly even in the absence of
some of the perfect adaptation conditions. In finding the perfect
adaptation conditions, we focus on studying equations which do not
depend on any individual methylation levels, these global
equations are obtained by either conservation laws or summing
steady-state equations over different methylation levels. The same
approach is also useful in understanding the near perfect
adaptation when the perfect adaptation conditions are violated.
Technically, we can explain the near perfect adaptation by the
existence of approximate global equations replacing the ones lost
due to the violation of perfect adaptation conditions.
Biologically, these approximate global equations are caused by
various intrinsic properties of the system, such as separation of
scales in protein concentrations and reaction rates, or specific
properties of the receptor distribution in different methylation
states. Since real biological systems are not likely to satisfy
all the perfect adaptation conditions exactly, the abundance of
such near perfect adaptation regions in the parameter space
strongly limits the range of activity variation and is probably
responsible for the robustness of the system's ability to adapt
almost perfectly.

Through systematic study of the system's behavior when different
perfect adaptation conditions are violated, we have also
identified parameter regions where significant deviation from
perfect adaptation occurs. This may provide possible explanations
to bacterial chemotaxis responses that does not adapt accurately,
such as the serine response as reported in (Berg, 1972), and
constitute concrete predictions that can be experimentally
verified.

Aside from perfect adaptation, another challenge for modelling
bacterial chemotaxis is to understand the large signal
amplification from ligand concentration change to the change in
bacterium flagella rotation bias. To directly compare between
experiments and simulation, detailed information between CheY-P
concentration and the motor rotation bias is needed. Recently, the
connection between CheY-P level and the motor activity was
investigated in (Scharf, 1998; Alon, 1998; Cluzel, 2000). In
(Cluzel, 2000), where rotation bias of single bacterium was
measured for different [Y$^P$] concentrations, it was shown that
the motor bias for individual bacterium should be fitted by a Hill
function with a large Hill coefficient $(\sim10)$. This highly
nonlinear function may explain the advantage of perfect adaptation
in amplifying the gain, and also the nonlinear dependence of
$B_\text{CCW}$, the CCW rotation bias, on changes in ligand
occupancy as found in (Jasuja, 1999). However, quantitatively,
from (Cluzel, 2000), the maximum signal amplification from change
in $[Y^P]$ to the tumbling frequency is measured to be:
$dB_\text{ccw}/d\ln[Y^P]\approx 2.2$. With the reference
parameters in our model, this leads to a total signal
amplification of $dB_\text{CCW}/d\ln[Y^P] \times d\ln[Y^P]/dL
\approx 2.2\times0.65\approx1.43$, which is still much too small
as compared with the total signal amplification measured in
experiments, e.g., $\sim 30$ as reported in (Jasuja, 1999).

The gain of the system could come from receptor clustering as
suggested in (Bray, 1998). However, to reconcile the existence of
high gain and the wide dynamic range of response, it is highly
desirable to have high gain for the signal transduction pathway
itself. One of the interesting findings of our study is that if
the system satisfies all the perfect adaptation conditions, the
steady state activity of the system is independent of the exact
values of the receptor activity $P\nl$ for $n\in [1,3]$. On the
other hand, the response of the system, { defined here as the
difference of CheY-P concentrations between its extreme value
after the stimulus and its original value before the stimulus},
directly depends on the difference of receptor activity between
ligand bound and ligand free receptors: $\Delta P_{n}\equiv
P_{nv}-P_{no}$. The higher these differences are, the higher the
response will be. In order to have high response, it is favorable
to increase $\Delta P_{n}$ and to have lower total activity.
Indeed, if we simply increase the activity difference between the
ligand bound and vacant receptor, such as those used in Fig.
\ref{Fig:condRB0} $c$, the total amplification can be increased
to: $2.2\times1.7=3.74$. Other changes, such as reducing the
system's total activity, can enhance the gain much more, as noted
also in (Barkai, 2001). A detailed study of the response of the
system is outside the scope of this paper and will be reported in
another communication.

Overall, the current model is capable of explaining the
qualitative behaviors of the chemotaxis pathway related to
adaptation, in particular, the robustness of the system's ability
to adapt nearly perfectly. Much work is still needed to modify and
enrich the model to understand the high sensitivity and wide
dynamic range of the system (Sourjik, 2002). { However, because
adaptation and response occur with very different time scale and
via largely different molecular processes, modification of the
model in explaining the high response gain should not change the
perfect adaptation conditions significantly. Indeed, it is not
hard to show that even with receptor coupling added to the current
model, the conditions we identified in this paper are still needed
for the system to achieve perfect adaptation, the only change is
that activity of each receptor now depends also on its neighbors'
activities (B. Mello and Y. Tu, manuscript in preparation). We
believe that, as long as the basic structure of the protein
interaction network stays intact, the perfect adaptation
conditions identified here will be mostly valid. These conditions
not only offer explanation for adaptation accuracy and its
robustness, furthermore, they serve as constrains for constructing
quantitative models in understanding other aspects of the
bacterial chemotaxis.}

\section*{APPENDIX}

In this section, we describe the detailed derivation of the
perfect adaptation conditions listed in the section ``Conditions
for perfect
adaptation".
 As described there, 
the approach is to construct global equations using
global and composite variables that do not depend on the receptor
population in any {\it one} individual methylation state.

First, we concentrate on the methylation related equations. Eq.
\ref{Bf1}, Eq. \ref{Bf2} and summation of Eq. \ref{Bf3} over
$n\in[0,4]$ gives 3 global equations. For the steady state, the
methylation flux between different methylation states should be
zero:
\begin{equation}
\label{JC}
 J_{n}=k^R_n\frac{[R^F][T^F_n]}{K^R_n}
     -   k^B_{n+1}\frac{[B^{PF}][T^F_{n+1}]}{K^B_{n+1}}=0,\;\;\;(0\le n\le
     3).
\end{equation}
Condition \ref{Cond:rkRkB} can be used in factoring out the common
$n$ dependent factor from $k^{R}_{n+1}$ and $k^{B}_n$ in $J_n$,
after which Eq. \ref{JC} are summed over $n\in[0,3]$ to obtain a
global equation.

Using condition \ref{Cond:KLinear}, the Michaelis constants can be
expressed as $K^R\nl=K^R/(P_{4\lambda}-P\nl)$ and
$K^B\nl=K^B/(P\nl-P_{0\lambda})$, where $K^R$ and $K^B$ are
constants. If we further enforce condition 3, i.e.,
$P_{4o}=P_{4v}\equiv P_4 $ and $P_{0o}=P_{0v}\equiv P_0$, we can
convert all the weighted sums of the individual receptor
concentrations into two composite receptor concentrations $[T^F]$
and $[T^{AF}]$. $[T^F] \equiv \sn[T^F_n]$ is the total
concentration of the free receptors; $[T^{AF}] \equiv \sn
P_n[T^F_n]$ is the total concentration of the active free
receptors, where $P_n\equiv(1-L_n)P_{nv}+L_nP_{no}$ is the
population-weighted average activity for a receptor with $n$
methyl groups. Therefore, after applying conditions 2, 3 and 4,
the 4 methylation related global equations can be written as:
\begin{gather}
  [R^T] = \label{RFR}
  [R^F](1+P_{4}\frac{[T^F]}{K^R}-\frac{[T^{AF}]}{K^R}),\\
  [B^{P}] =
  [B^{PF}](1-P_0\frac{[T^F]}{K^B}+\frac{[T^{AF}]}{K^B}),\\
\label{TT}
  [T^T]
  =\left(1+P_{4}\frac{[R^F]}{K^R}-P_0 \frac{[B^{PF}]}{K^B}\right)[T^F]\nonumber\\
      +\left(-\frac{[R^F]}{K^R}+\frac{[B^{PF}]}{K^B}\right)[T^{AF}],\\
 \label{SSR}
  k^R [R^F]\frac{\left(P_{4}[T^F]-[T^{AF}]\right)}{K^R}-\nonumber\\
  k^B [B^{PF}]\frac{\left(-P_0[T^F]+[T^{AF}]\right)}{K^B} = 0.
\end{gather}


If $K^B\nl$ is a constant (i.e., CheB-P binds equally to all
receptors), condition 2 is violated. However, it is not hard to
see that if the receptor population in the $n=0$ methylation
state, $[T_0]$, is small, we can still sum up the methylation
balance equations to form a global equation. The same is true if
$K^R\nl$ is a constant and $[T_4]\approx 0$.

Next, we focus on the phosphorylation related equations. Besides
its importance in producing the final output of the signal
transduction pathway CheY-P, the phosphorylation is also coupled
back to the methylation process through concentration $[B^{PF}]$.
By writing $k^P\nl \equiv k^PP\nl$ and using condition
\ref{Cond:BYP2}: $k^{PY}\nl \equiv k^{PY}P\nl$ and $k^{PB}\nl
\equiv k^{PB}P\nl$, the phosphorylation related global equations
can be written as:
\begin{gather} \label{YPR}
  [Y^P] = \frac{k^{PY}[T^{PA}]}{k^{HY}+k^{PY}[T^{PA}]}[Y^T],\\
  [B^{PF}] = \frac{k^{PB}[T^{PA}]}{k^{HB}+k^{PB}[T^{PA}]}[B^{F}]
  \label{BPFR},\\
 \label{TPAR}
  [T^{PA}]
        =    \frac{[T^{A}]}{1+\frac{k^{PY}}{k^P}([Y^T]-[Y^P])
               +\frac{k^{PB}}{k^P}([B^{F}]-[B^{PF}])},
\end{gather}
Eq. \ref{TPAR} is obtained by summing Eq. \ref{YP3} over
$n\in[0,4]$. There are two composite variables, $[T^PA]$ and
$[T^A]$ in the above equations. $[T^{A}]\equiv \sn P_n[T_n]$ is
the total concentration of active receptors, $[T^{PA}] \equiv \snl
P\nl[T^{P}\nl]$ is phosphorylated active receptor concentrations.

If the CheA phosphate transfer rates are independent of its
ligand/methylation status, i.e., $k^{PY}\nl \equiv k^{PY}$ and
$k^{PB}\nl \equiv k^{PB}$, condition 5 is broken. A new composite
variable $[T^P]\equiv \sn [T_n^P]$ appears in the above equations,
replacing $[T^{PA}]$ in Eq. \ref{YPR}, Eq. \ref{BPFR} and part of
Eq. \ref{TPAR}. However, if $[T^{PA}]\ll [T^{A}]$, e.g., due to
efficient phosphate transfer from CheA to CheY, $[T^{PA}]$ can be
neglected, and there again will be only two composite variables in
the phosphorylation related global equations, and therefore the
system may still adapt near perfectly in absence of condition 5,
as discussed in the section ``Violating condition 5".

The methylation and the phosphorylation global equations
communicate through various CheB concentrations. An extra equation
is necessary to connect the concentrations of these different
forms of the same proteins:
\begin{equation} \label{BTR1}
  [B^T] = [B^{P}] + [B^{F}] - [B^{PF}].
\end{equation}

Finally, by using Eq. \ref{Bf3}, we can write down the expression
for the total receptor activity of the system $[T^A]$ that appears
in Eq. \ref{TPAR}:
\begin{eqnarray}
  [T^{A}]&=&\sn P_n[T_n]\nonumber\\ \label{TATR}
  &=&\left(1+P_{4}\frac{[R^F]}{K^R}-P_0\frac{[B^{PF}]}{K^B}\right)[T^{AF}]\nonumber\\
  &+& \left(-\frac{[R^F]}{K^R}+\frac{[B^{PF}]}{K^B}\right)
  \sn {P_n}^2[T^F_n].
\end{eqnarray}
The above equation contains a new composite variable $\xi'=\sn
{P_n}^2[T^F_n]$. Condition \ref{Cond:rRB} is thus required to
eliminate this extra term. Part of $\xi'$ can be expressed in
terms of the other composite variables, such as $[T^F]$ and
$[T^{AF}]$. Therefore, the effect of violating condition 6 can not
be simply measured by the value of $\xi'$, as we discussed in the
section
``Violating condition 6".

If all the conditions listed in Conditions for perfect adaptation 
are satisfied, we have
nine global equations: Eqs. \ref{RFR}-\ref{TATR}, these 9 global
equations contains 5 global variables: $[R^F]$, $[B^{P}]$,
$[B^{PF}]$, $[B^{F}]$, $[Y^P]$, and 4 composite variables:
$[T^F]$, $[T^{AF}]$, $[T^{A}]$, $[T^{PA}]$. Therefore, the steady
state values of all the nine global or composite variables,
including $[Y^P]$, will be independent of the ligand concentration
and the system can achieve perfect adaptation.

We are thankful to Jeremy Rice, Geofferey Grinstein and Gustavo
Stolovitzky for helpful discussions and careful reading of the
manuscript. B. Mello has a scholarship from CNPq -- Brazil.

\section*{REFERENCES}
\begin{description}

\item
Alon,~U., L.~Camarena, M.G. Surette, B.~Aguera y~Arcas, Yi~Liu, S.~Leibler, and
  J.~B. Stock.
\newblock 1998.
\newblock Response regulator output in bacterical chemotaxis.
\newblock \textit{ EMBO J.} 17:4238--4248.

\item
Alon,~U., M.G. Surette, N.~Barkai, and S.~Leibler.
\newblock 1999.
\newblock Robustness in bacterial chemotaxis.
\newblock \textit{ Nature} 397:168--171.

\item
Asakura,~S., and H.~Honda
\newblock 1984.
\newblock Two-state model for bacterial chemoreceptor proteins.
The role of multiple methylation.
\newblock \textit{ J. Mol. Biol.} 176:349-367.

\item
Barkai,~N., and S.~Leibler.
\newblock 1997.
\newblock Robustness in simple biochemical networks.
\newblock \textit{ Nature} 387:913--917.

\item Barkai, N., U. Alon, and S. Leibler.
\newblock 2001.
\newblock Robust amplification in adaptive signal transduction networks.
\newblock \textit{ C. R. Acad. Sci. Paris} 2:1--7.

\item
Barnakov,~A.N., L.A. Barnakova, and G.L. Hazelbauer.
\newblock 1999.
\newblock Efficient adaptational demithylation of chemoreceptors requires the
  same enzyme-docking site as efficent methylation.
\newblock \textit{ Proc. Natl. Acd. Sci. USA} 96:10667--10672.

\item
Berg,~H.C., and D.A. Brown.
\newblock 1972.
\newblock Chemotaxis in Escherichia coli analysed by
Three-dimensional Tracking
\newblock \textit{ Nature} 239:500--504.

\item
Borkovich,~K.A., L.A. Alex, and M.I. Simon.
\newblock 1992.
\newblock Attenuation of sensory receptor signaling by covalent
modification.
\newblock \textit{ Proc. Natl. Acad. Sci. USA} 89:6756-6760.

\item
Bornhorst, J.A., and J.J. Falke
\newblock 2001.
\newblock Evidence that Both Ligand Binding and Covalent
Adaptation Drive a Two-state Equilibrium in the Aspartate Receptor
Signaling Complex.
\newblock \textit{ J. Gen. Physiol.} 118:693-710.

\item
Bourret,~R.B., and A.M. Stock
\newblock 2002.
\newblock Molecular information processing: lessons from bacterial
chemotaxis
\newblock \textit{ J. Biol. Chem.} 277:9625--9628.

\item
Bray,~D., R.B. Bourret, and M.I. Simon.
\newblock 1993.
\newblock Computer simulation of the phosphorylation cascade controlling
  bacterial chenmotaxis.
\newblock \textit{ Mol. Bio. Cell} 4:469--482.

\item
Bray,~D., M.D. Levin, and C.J. Morton-Firth.
\newblock 1998.
\newblock Receptor clustering as a cellular mechanism to control sensitivity.
\newblock \textit{ Nature} 393:85--88.

\item
Bren,~A., and M.~Eisenbach.
\newblock 2000.
\newblock How signals are heard during bacterial chemotaxis: protein-protein
  interactions in sensory signal propagation.
\newblock \textit{ J. Bacteriol.} 182:6865--6873.

\item
Cluzel,~P., M.~Suette, and S.~Leibler.
\newblock 2000.
\newblock An ultrasensitive bacterial motor revealed by monitoring signaling
  proteins in single cells.
\newblock \textit{ Science} 287:1652--1655.

\item
Djordjevic,~S., P.N. Goudreau, Qingping Xu, A.M. Stock, and A.H. West.
\newblock 1998.
\newblock Structural basis for methylesterase CheB regulation by a
  phosphorylation-activated domain.
\newblock \textit{ Proc. Natl. Acad. Sci. USA} 95:1381--1386.

\item
Dunten,~P., and D.E. Koshland, Jr.
\newblock 1991.
\newblock Tuning the responsiveness of a sensory receptor via
covalent modofication
\newblock \textit{ J. Biol. Chem.} 266:1491--1496.

\item
Falke,~J.~J., R.B. Bass, S.L. Butler, S.A. Chervitz, and N.A. Danielson.
\newblock 1997.
\newblock The two-component signaling pathwaty of bacterial chemotaxis.
\newblock \textit{ Annu. Rev. Cell Dev. Biol.} 13:457--512.

\item Hauri,~D.C., and J. Ross.
\newblock 1995.
\newblock A model of excitation and adaptation in bacterial chemotaxis.
\newblock \textit{ Biophys. J.} 68:708--722.

\item
Jasuja,~R., Yu-Lin, D.R. Trentham, and S.~Khan.
\newblock 1999.
\newblock Response tuning in bacterial chemotaxis.
\newblock \textit{ Proc. Natl. Acad. Sci. USA} 96:11346--11351.

\item
Liu,~Y., M.~Levit, R.~Lurz, M.G. Surette, and J.B. Stock.
\newblock 1997.
\newblock Receptor-mediated protein kinase activation and the mechanism of
  transmembrane signaling in bacterial chemotaxis.
\newblock \textit{ EMBO J.} 16:7231--7240.

\item%
Lukat, G.S., B.H. Lee, J.M. Mottonen, A.M. Stock, and J.B. Stock.
\newblock 1991.
\newblock Roles of the high conserved aspartate and lysine
residues in the response regulator of bacterial chemotaxis.
\newblock \textit{ J. Biol. Chem.}, 266:8348-8354.

\item
Morton-Firth,~C.J., and D.~Bray.
\newblock 1998.
\newblock Predicting temporal fluctuations in an intracellular signalling
  pathway.
\newblock \textit{ J. Theor. Biol.} 192:117--128.

\item
Morton-Firth,~C.J., T.S. Shimizu, and D.~Bray.
\newblock 1999.
\newblock A free-energy-based stochastic simulation of the tar receptor
  complex.
\newblock \textit{ J. Mol. Biol.} 286:1059--1074.

\item
Othmer, H.G. and P. Schaap.
\newblock 1998.
\newblock Oscillatory cAMP signaling in the development of Dictyostelium discoideum.
\newblock \textit{ Comments on Theoretical Biology}, 5:175--282.

\item
Scharf,~B.E., K.A. Fahrner, L.~Turner, and H.C. Berg.
\newblock 1998.
\newblock Control of direction of flagellar rotation in bacterial chemotaxis.
\newblock \textit{ Proc. Natl. Acad. Sci. USA} 95:201--206.

\item
Shapiro,~M.J., and D.E.~Koshland, Jr.
\newblock 1994.
\newblock Mutagenic studies of the interaction between the aspartate receptor
  and methyltransferase from escherichia coli.
\newblock \textit{ J. Biol. Chem.} 269:11054--11059.

\item
Shapiro,~M.J., D.~Panomitros, and D.E.~Koshland, Jr.
\newblock 1995.
\newblock Interactions between the methylation sites of escherichia coli
  aspartate receptor mediated by the methyltransferase.
\newblock \textit{ J. Biol. Chem.} 270:751--755.

\item
Sourjik,~V., and H.C. Berg.
\newblock 2002.
\newblock Receptor sensitivity in bacterial chemotaxis.
\newblock \textit{ Proc. Natl. Acad. Sci. USA} 99:123--127.

\item
Spiro,~P.A., J.S. Parkinson, and H.G. Othmer.
\newblock 1997.
\newblock A model of excitation and adaptation in bacterial chemotaxis.
\newblock \textit{ Proc. Natl. Acad. Sci. USA} 94:7263--7268.

\item
Yi,~T., Yun Huang, M.I. Simon, and J.~Doyle.
\newblock 2000.
\newblock Robust perfect adaptation in bacterial chemotaxis through integral
  feedback control.
\newblock \textit{ Proc. Natl. Acad. Sci. USA} 97:4649--4653.

\end{description}

\end{document}